# Optimizing Sensor Node Localization for Achieving Sustainable Smart Agriculture System Connectivity

[1*]Mohamed Naeem

[1]Computer Networks and Data Centre- Cairo,

Arab Academy for Science, Technology and Maritime Transport, Cairo 11799, Egypt,

mohamed.hamdy@aast.edu

**Abstract**

The innovative agriculture system is revolutionizing how we farm, making it one of the most critical innovations of our time! Yet it faces significant connectivity challenges, particularly with the sensors that power this technology. An efficient sensor deployment solution is still required to maximize the network's detection capabilities and efficiency while minimizing resource consumption and operational costs. This paper introduces an innovative sensor allocation optimization method that employs a Gradient-Based Iteration with Lagrange. The proposed method successfully enhances coverage by using this hybrid approach while minimizing the number of sensor nodes required in a grid-based allocation. The proposed sensor distribution outperformed the classic deterministic deployment across coverage, number of sensors, cost, and power consumption. Furthermore, scalability is enhanced by extending sensing coverage to the remaining area using Bluetooth technology with a shorter communication range. Moreover, the proposed algorithm achieved 98.5% wireless sensor coverage, compared with 95% for the particle swarm distribution.

1. Introduction

The smart agriculture system has emerged as one of the most intriguing innovations in the agricultural sector, primarily aimed at optimizing the food supply chain [1]. This system encompasses various forms and components tailored to specific agricultural environments and applications. However, certain elements, such as sensors and actuators, remain consistent across different implementations. Sensors serve as data acquisition tools, while actuators execute decisions based on the data they acquire. Both components require robust security, reliable connectivity, efficient power management, and regular maintenance [2]. One primary challenge in smart agriculture is ensuring reliable connectivity, which can significantly influence other factors, such as power usage and overall costs. Connectivity issues can arise from multiple factors, including the type of communication technology used, the location's geographical features, the layout of the agricultural fields, and environmental parameters [3].

Various communication technologies [4] are employed in smart agriculture systems, including wireless, optical, satellite, and wired technologies. While these technologies have demonstrated varying effectiveness in implementing smart agriculture solutions, wireless connectivity is the most deployable one [5]. The geographic location of agricultural areas often requires the use of standard technologies such as satellite connectivity, NB-IoT, Sigfox, LoRa-WAN, and Cat-M1. Additionally, strategies to address location constraints include implementing smart agricultural management systems as an Artificial Intelligence Internet of Things (AIoT) solution that leverages edge computing [6]. To meet connectivity expectations, field owners can be assured of cellular data coverage in certain regions through technologies such as 5G, 4G, and 3G. Most agricultural zones require wireless connectivity within industrial field parameters, favoring low-range wireless solutions. This type of connectivity is essential for ensuring robust communication among smart agricultural components, particularly actuators and sensors [7]. Wireless technologies commonly used for these applications include Zigbee, WiFi, Bluetooth, Z-Wave, and LoRa [8]. To enhance connectivity with these technologies, it is crucial to position system sensors strategically throughout the agricultural area. Thoughtful sensor placement reduces power consumption and latency. Minimizing the number of sensors in large agrarian fields is important while preserving effective field-sensing capabilities [9]. Furthermore, maintaining interoperability among sensors and actuators is essential [10]. All these factors underscore the importance of optimal sensor allocation within agricultural fields. There are diverse solutions that tackle those factors individually. To our knowledge, there's no definitive solution in the literature that addresses all of those issues simultaneously.

This paper presents an optimized sensor allocation strategy to enhance sensing within the framework of sustainable connectivity among elements of smart agricultural systems. The structure of the paper includes a related work section that reviews recent efforts in the research field, a Methodology section that outlines the strategy employed for conducting the research and obtaining results, a section on Simulation and results that discusses the simulations used to evaluate the proposed solution and provides a detailed analysis of the sensor allocation method, and a conclusion that summarizes the research approach, findings, and suggestions for future work.

## 2. Related Work

Numerous related studies in the literature address the sensor allocation problem to enhance the sustainability of smart agriculture system connectivity, particularly for sensors and actuators. These efforts can be categorized into classical, metaheuristic, and self-optimization techniques. This research focuses on the allocation of stationary sensor nodes, making classical allocation optimization methods particularly relevant. A force-based technique, VFA, has been documented in [11]. An EVFA force-based allocation technique has also been reported in [12]. Other methods include using the Van der Waals force as a basis for allocation, as mentioned in [13].

Furthermore, a computational geometry approach to sensor allocation using the Delaunay triangulator has been explored in [14], and the Voronoi diagram in [15]. On the classical front, one of the most intriguing methods involves dividing the agricultural area into a square grid. Sensor allocation is then performed based on this grid distribution, with total area optimization achieved through triangular grids [16], square grids [17], or hexagonal grids [18]. However, the uncertainty inherent in sensor allocation within the grid system creates a research gap, prompting the exploration of new optimization methods. This study employs a classical grid distribution algorithm and evaluates it against conventional grid distribution, Gaussian methods [19], and the PSO method [20] while also proposing a novel optimization algorithm.

## 3. Network Model

Table 1 presents common notations used in our analysis. We consider a two-dimensional sensing field, $A = X \times Y$, divided into a square grid of equal-sized cells. The relationship between these cells and the targets is represented as a bipartite graph (A, T), where A denotes the set of cells, and T signifies the set of fixed targets. Those fixed targets represent the plant tree under monitoring.

For instance, an edge c → t, where a ∈ A and t ∈ T, implies that a node in cell c can monitor target t. Each cell c is characterized by its coordinates a (x, y), with x = 1,..., X and y = 1,..., Y. Let N represent the set of sensor nodes, and we use Ni and Tj to index the sensor nodes and targets, respectively.

Table 1: Agricultural model parameters

| Symbol | Description |
| --- | --- |
| A(X, Y) | Agricultural area as a dimensional description of the horizontal X dimension and the Vertical Y dimension |
| (A, T) | The agricultural field is defined as an area A containing a set of cells and target trees T. |
| c | Cell or unit area of the total agricultural area A. |
| t | Single objects are tracked or sensed from the total T objects. |
| a | Sub area of the total agricultural area with a small horizontal distance x, and a small vertical distance y. |
| Ni | Ith sensor node |
| Tj | Jth tracked or sensed object. |
| T(ni) | Sensing coverage area of the ith sensor node for a target T |
| C(tj) | Cells contain the jth target that is being tracked or sensed. |
| W(x,y) | Weight vector of a cell containing a sensor node |
| P(x,y) | Power consumption of a sub-area a(x,y) containing sensor nodes |
| R | The wireless propagation from the sensor node antenna is assumed to be ideal, omnidirectional, and circular. |
| di | The ith distance of the ith sensor node from the anchor node, with I is a counter 1,2,3,4,…N |

All targets are assumed to be stationary and evenly spaced within each cell, with a 5-meter interspacing. However, there is uncertainty for the number of sensing nodes and the location of each one. Let $T(n_i)$ represent a function that returns the set of targets covered by sensor node $N_i$. Conversely, the function $N(t_j)$ provides the set of sensor nodes that cover target $t_j$. To simplify notation, we use T(x,y) to denote the set of targets covered by cell a(x,y). If a sensor node ni is placed in cell a(x,y), it will cover $T(n_i)$ targets. Let $C(t_j)$ represent the cells covering target $t_j$. The weight of a cell is defined as w(x,y) = P(x,y) × Tc(x,y), reflecting both the cell's consumption rate and the number of targets that could be covered if a sensor node were placed in that cell. The sensing node coverage is shown in Figure 1.

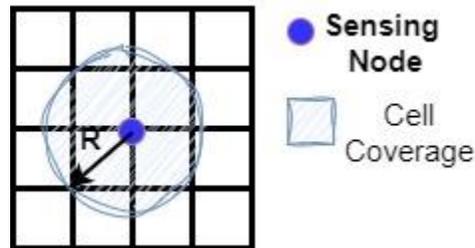

*Figure 1 Sensing node cell coverage*

The sensing node is assumed to be an isotropic wireless radiator with a circular radiation pattern, modeled as an ideal radiator of radius R. The node's wireless coverage spans several square cells, as illustrated in Figure 1, through which the node is active. Sensor nodes can cover up to β cells.

This paper assumes that the sensor nodes are equipped with omnidirectional sensors, including both soil and weather sensors. For instance, in Figure 1, we observe a sensor node $n_i$ situated in cell c (3,3), with targets located in cells A (2,3) and A (4,4). The dashed circle indicates the sensing range of sensor node $n_i$, as shown by the blue cells. As illustrated in Figure 1, the sensing range of node $n_i$ encompasses β = four cells. A more accurate approximation of the sensing range can be obtained by using smaller cell sizes. We now formulate the sensor N-node placement problem as an integer linear program (ILP). Let N (x, y) be the number of sensor nodes that cover cell A (x, y). Recall that the energy consumption rate in cell A (x, y) is P (x, y). The ILP aims to achieve a minimum number of sensor nodes while maximizing coverage and minimizing power consumption.

$$\sum N(x, y) \quad (x, y) \in A \quad (1)$$

**Where:**

$$A_{x,yJ\in(\tau_j)}^{\sum P(x,y) x N}(x, y) \geq P, \forall T_j \in T \quad (2)$$

$$N(x, y) \in T + \forall (x, y) \in A \quad (3)$$

Constraint (2) guarantees that the sensor nodes can function continuously over time while successfully collecting the necessary number of samples from the designated targets. This ensures the sensor network's longevity and efficiency. Meanwhile, Constraint (3) requires that the number of sensor nodes deployed in each cell be a non-negative integer, ensuring that only whole units of sensor nodes can be allocated and preventing fractional assignments that could compromise operational integrity.

In addition to this, the deployment of sensor nodes depends on one of three techniques:
- Random sensor allocation: Stochastic deployment of the sensor nodes through the application area.
- Statistical sensor allocation: Probabilistic methods used to deploy the sensor node while considering target sensing and process flow.
- Deterministic sensor allocation: Systematic methods used to deploy sensors by considering a referencing point, object, or central unit.

This research adopts a deterministic approach, allocating the central sensor units at the center of the area. Followed by allocating the sensor nodes with an equal distance from the central sensor node as given in equation (4).

$$D_i = d_i * \cos(\psi) \quad (4)$$

Where:
- i is the number of sensor nodes distributed across the agricultural area,
- ψ is the angle between the horizontal line through the central sensor node and the line from the central node to the sensor node's allocation point Ni.

Given the constraints in equations 2 and 3, the research adopts a clustering-based deployment. The clustering is based on deterministic area division, that is, dividing the overall agricultural area into sub-areas. The division is governed by equation (5).

$$A_c = (L * M) * (L * M) \text{ m}^2 \quad (5)$$

Where:
- M is the maximum repetition or recurring factor
- L is the maximum deployment distance.

Both L and M depend on the wireless technologies employed within the solution. If the solution uses Wi-Fi, then M = 4, the maximum possible repetition of the Wi-Fi signal, including the source. Also, L = 50 meters, the maximum possible deployment distance over which the Wi-Fi signal can propagate [21]. The cluster area, as a sub-agricultural area within the total agricultural area, and the wireless technology considered are determined by $A_c(x, y) = A_c(300, 300)$.

## 4. Methodology

The methodology Starts by dividing the $A(x, y) = (300, 300)$ m agricultural area into a systematic grid, as shown in Figure 2. The systematic grid contains 36 square cells, each with an edge length of 50 meters. This division aims to solve the ILP problem over a sample area $(x_s, y_s) = (50, 50)$, comprising 36 cells and 3600 targets, as illustrated in Figure 4. While the targets are systematic tree plants, the solution is optimized into the systematic grid numbers. This solution guarantees better coverage and fewer sensor node allocations, effectively contributing to achieving the minimum perpetual coverage node placement (MPCNP) [22].

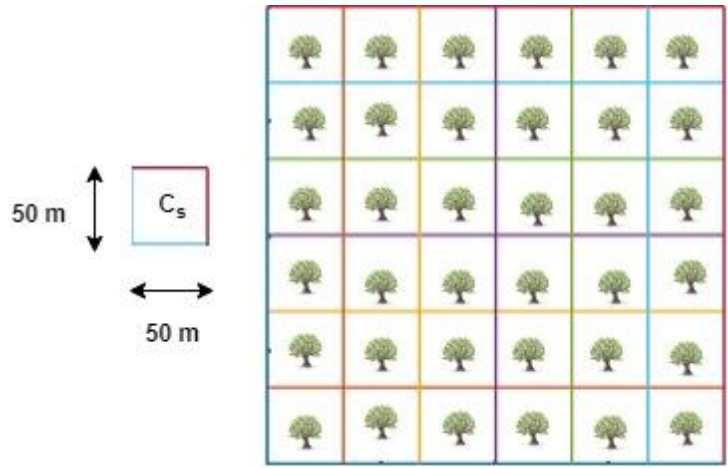

Figure 2: Systematic grid distributed area

While assuming a sensor distribution with an omnidirectional antenna pattern, the coverage and power consumption require careful wireless infrastructure design on one side and a well-planned allocation on the other. For those reasons, the methodology aims to achieve efficient heterogeneous wireless coverage by utilizing decision-making algorithms based on quantitative fuzzy systems. This aims to rank wireless networks based on range, power consumption, and delay constraints. However, improving node allocation can be achieved by using minimum-square cells.

### 4.1 Heterogeneous Wireless Infrastructure Method

The Fuzzy Analytical Hierarchy Process (FAHP) assesses and ranks heterogeneous wireless networks as a sustainable wireless network infrastructure for sensors and actuators. The fuzzy module quantitatively assesses the four selected wireless networks, as illustrated in Figure 3. Then, it ranks them based on the four connection parameters, delay, power consumption, range, and bandwidth, in a hierarchical relationship illustrated in Figure 4 [23].

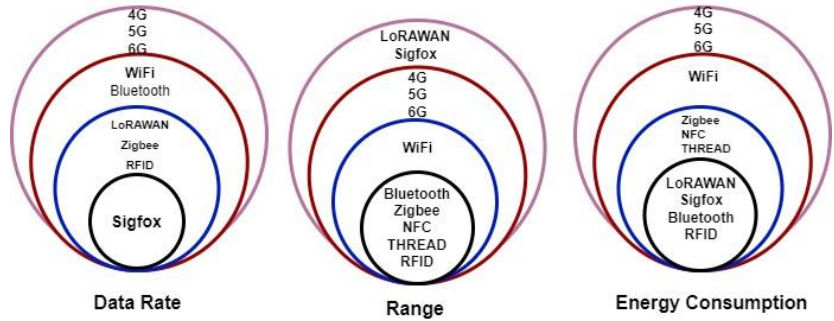

Figure 3 Wireless Communication Networks

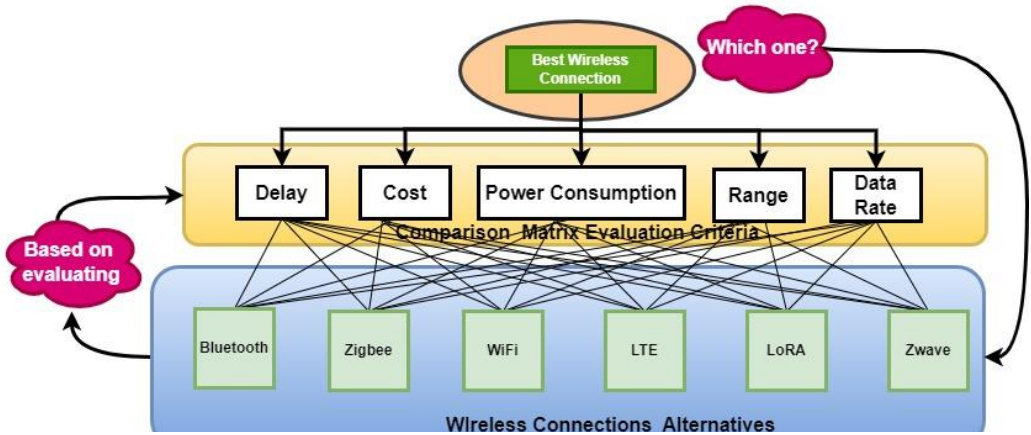

Figure 4 Wireless connectivity selection hierarchical relationship

As shown in Figure 4, wireless communication networks exhibit a wide range of data rates, power consumption, and coverage ranges. This diversity creates uncertainty in selecting and prioritizing them as the best wireless communication infrastructure for sensor networks. A fuzzy algorithm provides a quantitative assessment of the pairwise comparison matrix by establishing a hierarchy of relationships, moving away from the traditional expert-driven evaluations used in the original AHP process, as illustrated in Figure 4. Subsequently, this comparison matrix is normalized and applied within the AHP subprocesses to rank the wireless connectivity options, as demonstrated in Figure 5.

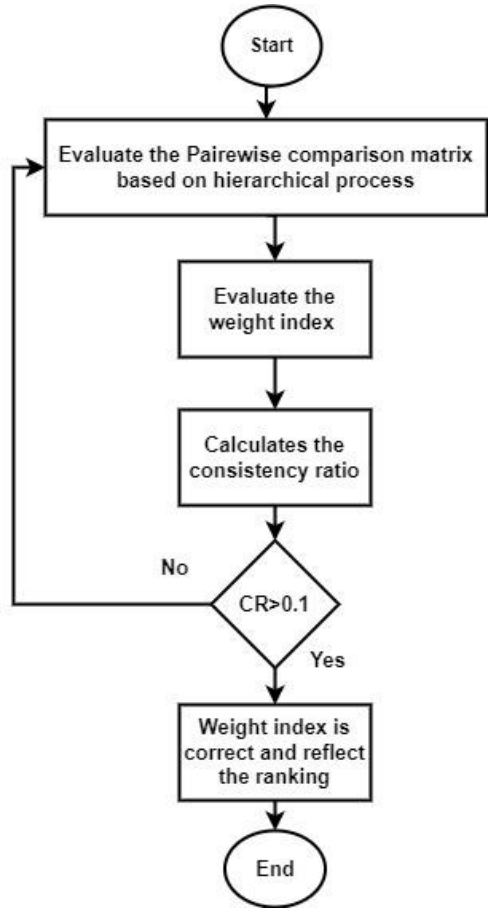

Figure 5: Heterogeneous wireless infrastructure selection

4.2 Hybrid method for sensor node allocation

The adopted hybrid sensor allocation method comprises systematic grid division of the agricultural area, followed by mathematical allocation and optimization. The methodology assumes the following for the model considered:

- All sensors are static and fixed
- The sensor nodes are heterogeneous in function and connectivity
- The total area is divided into sub-areas as a cluster, and each one has a cluster controller
- The sensor nodes' distribution depends on a controller node at the center of the agricultural area
- All sensor nodes propagate a circular radiation pattern (using an ideal omnidirectional antenna)
- The cluster controllers are connected to a central controller that manages the total area.

Since sensor nodes are heterogeneous in function, it's preferable to state this as shown in Table 2.

Table 2 Sensor node categories

| Sensor node Type | Features |
|---|---|
| Normal sensor node | Sensing of the target objects only |
| Anchor sensor node | Management of cluster nodes and sensing of the target objects. |

The allocation of sensor nodes to effectively sense agricultural objects while minimizing deployment units is achievable through systematic grid distribution, in which a node is placed within a selected grid cell. One of the most effective optimization methods is the Fibonacci series, which can be used to select the square grid sequence within the series. The Fibonacci method is a series of numbers that may be used to index the grid cell at (X, Y), as illustrated in equation (4).

$$F_n = F_{n-1} + F_{n-2} \qquad (6)$$

By solving equation 4, the Fibonacci method yields the minimum number of square grids required to enclose a sensor node within its perimeter. While Fibonacci offers a low-density grid deployment, it fails to provide adequate coverage. Additionally, the sensing ranges of nearby sensors may overlap. A random distribution of the same number of sensors effectively addresses this issue. Moreover, it's required to determine the location of the sensor nodes within the square grids. The most famous mathematical allocation technique is the old Trilateration, which is widely used to determine the distance of a sensor node from the assigned anchor node, as shown in Figure 6.

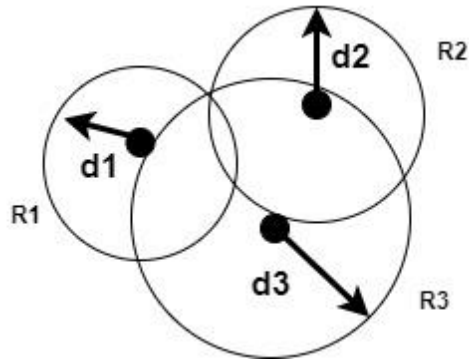

Figure 6 Trilateration

As shown in Figure 6, there are two undefined sensors with circular sensing coverage, with radii R1 and R2, and a defined central region R3. The allocation of those sensors, considering the anchor node (R3) as the central or reference node, is computed using equation 5 to calculate the di value.

$$d_i = \sqrt{((xi - X)^2 + (yi - Y)^2)} \quad (5)$$

When I refer to the individual sensor node number to be allocated relative to the anchor node, these distances are used to assign the sensor nodes. However, this method is efficient at allocating sensor nodes, but it is still not considered a tie, as there are many ways to allocate the nodes with several angles and alignment possibilities. To maintain optimal allocation while considering a more equitable distribution across the agricultural area, this distance should be measured relative to each node as the distance to the anchor node. The distance relationship given in equation (4) should satisfy the systematic distribution given in equation (5):

$$D_i = Q \times D_{i-1} \quad (6)$$

Where Q denotes the unit used to align sensor nodes with anchor nodes sequentially, optimizing node allocation should also ensure that the total number of nodes is no greater than the number obtained by the Fibonacci method. Numerous sensor node allocations optimize deployment distance [24]. While gradient descent yields more efficient results for determining the minimum distance, as in [25], it is less accurate when there are multiple constraints. Therefore, it's preferable to use a hybrid gradient descent algorithm that accounts for various constraints and optimizes the allocation to minimize distance. The proposed hybrid algorithm employs a gradient-based iteration with Lagrangian optimization to maintain a Fibonacci number of sensor nodes as a cost function, and a Lagrangian network model with constraints to achieve optimal wireless coverage while minimizing power consumption and delay.

The goal of sensor allocation is to adequately place a limited number of sensors to achieve the best possible performance for a given objective (maximizing coverage, minimizing placement uncertainty, maximizing detection probability, and providing minimum system delay) under certain constraints (overall cost, power consumption, and number of sensors). This is illustrated with the flowchart shown in Figure 8. Utilizing Fibonacci as the objective function to define the number of sensors. The constraints given in equations 2 and 3 should be considered in the determination of the inclined distance. The methodology then formulates the constrained optimization problems as shown in equation 1. Handling the constraints imposed by equations 2 and 3 to maximize sensor node availability, as given in equation 1, yields the Lagrangian in equation 8.

$$D(N1,\ldots,Nn,\lambda,\mu) = J(N1,\ldots Nn) - \sum_{i=1}^{m} \lambda_i \; Ti(N1,\ldots,Nn) - \sum_{j=1}^{u} \mu_j \; Pj(N1,\ldots,Nn) \quad (8)$$

Where λ is the Lagrangian multiplier for inequality constraints, and µ is the Lagrangian multiplier for equality constraints. Now, you can use a gradient-based iterative method to find the optimal sensor locations and the corresponding Lagrange multipliers. The approach is to initialize with an initial guess for the sensor locations and Lagrangian multipliers. Then iterate: for each iteration, compute the gradient of the Lagrangian with respect to the sensor location, and update the sensor location using gradient descent to minimize power consumption (P). Then update the Lagrangian multipliers and check for convergence after completing the iteration; convergence may occur due to either a change in sensor location below the threshold or to the constraints being satisfied with a very small tolerance, e.g., 0.01. The full process is shown in sequence in Figure 7.

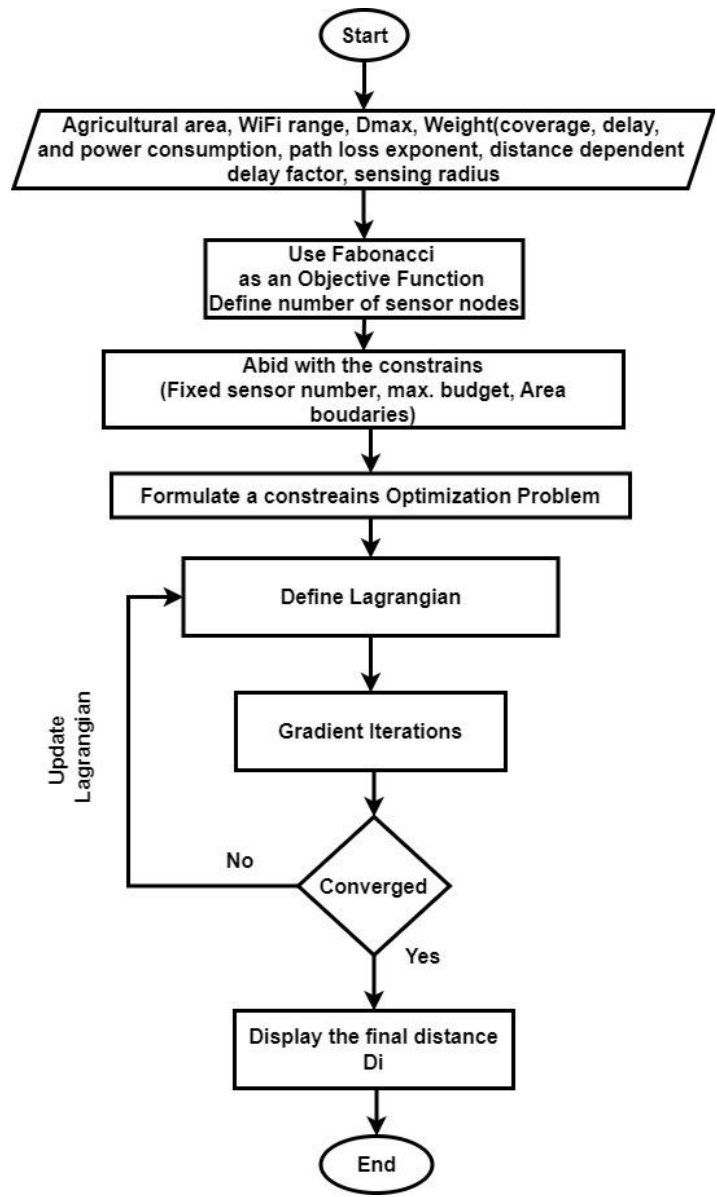

Figure 7: Flow chart of the proposed hybrid method

The pseudocode for determining the optimal inclined distance (di) and angle (PSi_i) for N sensor nodes relative to a central anchor, accounting for WiFi, power consumption, coverage, and delay, is shown in Figure 8.

```
// Inputs:
//   N: Number of sensor nodes
//   D_max: Maximum WiFi communication range (slant distance)
//   R_deploy_horizontal: Maximum horizontal deployment radius (optional)
//   Z_max_vertical: Maximum vertical deployment range (optional)
//   w_c, w_p, w_d: Weights for coverage, power, and delay
//   alpha: Path loss exponent for power consumption
//   kp: Power constant
//   kd: Base delay
//   cd: Distance-dependent delay factor
//   rs: Sensing radius of each sensor (in 3D space)
// Outputs:
//   optimal_distances: Array of N optimal inclined distances
//   optimal_angles_psi: Array of N optimal inclination angles
distances[1...N] = initialize_distances(N, D_max)
angles_psi[1...N] = initialize_angles(N) // Initialize between -pi/2 and pi/2
lambda_d[1...N] = 0   // For d_i >= 0
mu_d[1...N] = 0       // For d_i <= D_max
lambda_psi[1...N] = 0 // For -pi/2 <= psi_i <= pi/2 (can be handled as two inequalities)
mu_psi_upper[1...N] = 0
mu_psi_lower[1...N] = 0
gamma = 0             // For sum(1) = N
alpha_d_lr = 0.01
while iteration < max_iterations:
    iteration = iteration + 1
    current_coverage = calculate_coverage_3D(distances, angles_psi, rs, R_deploy_horizontal, Z_max_vertical)
    for i from 1 to N:
        gradients_d[i] = w_c * d_coverage_ddi - d_power_ddi - d_delay_ddi + lambda_d[i] - mu_d[i]
        d_coverage_dpsi_i = calculate_coverage_gradient_psi(distances, angles_psi, rs, R_deploy_horizontal, Z_max_vertical, i)
    new_distances[1...N] = 0
    for i from 1 to N:
        new_distances[i] = distances[i] + alpha_d_lr * gradients_d[i]
        new_distances[i] = max(0, new_distances[i])
        new_distances[i] = min(D_max, new_distances[i])
    distances = new_distances
        new_angles_psi[1...N] = 0
    for i from 1 to N:
        new_angles_psi[i] = angles_psi[i] + alpha_psi_lr * gradients_psi[i]
        new_angles_psi[i] = max(-pi / 2, new_angles_psi[i])
        new_angles_psi[i] = min(pi / 2, new_angles_psi[i])
        lambda_psi[i] = max(0, lambda_psi[i] - beta_lr * (-angles_psi[i] - pi / 2)) // For psi_i >= -pi/2
        mu_psi_upper[i] = max(0, mu_psi_upper[i] - beta_lr * (angles_psi[i] - pi / 2)) // For psi_i <= pi/2
    gamma = gamma - delta_lr * (N - N)
// Return the optimized distances and angles
optimal_distances = distances
optimal_angles_psi = angles_psi
```

Figure 8: Pseudocode of proposed algorithm

The code is designed to allocate the maximum distance to the sensor node, ensuring the cluster's node coverage, power consumption, delay, and capacity. By expanding the selected grids by a separation distance determined by the algorithm illustrated in Figure 2, we find that this distance equals the golden ratio, 1.618, thereby achieving improved optimization. This approach is mathematically validated to meet the constraints outlined in equation (2). Integrating both the Fibonacci sequence and the golden ratio yields a hybrid algorithm that can feasibly satisfy equations (2) and (3), which will be assessed through Simulation.

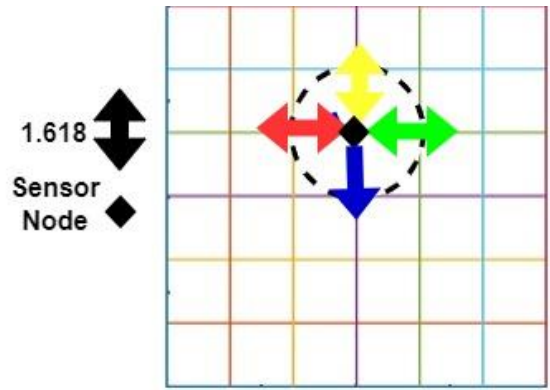

Figure 9: Illustration of sensor allocation spacing distance

The arrangement of the sensor nodes is determined by the cell grid, Fibonacci cell numbers, and the spacing guidance obtained from the gradient descent with the Lagrange method, as shown in Figure 9. The proposed solution begins by specifying an edge distance from the top of the area and placing the sensor nodes at the center of each cell. This deployment seeks to achieve better allocation of the sensor node through area A, as given in equation (1), along with satisfying the ILP problem containing power consumption and coverage, as presented in equations (2) and (3).

## 5. Simulation and Results

The MATLAB-based simulator used to analyze the ranking of heterogeneous wireless networks and to conduct forecasts using multiple linear regression is running on a Dell Latitude equipped with a 2.45 GHz Core i7 processor, 16 GB of RAM, and a 250 GB SSD.

5.1 Heterogeneous Wireless Communication Infrastructure

An evaluation and ranking of the six most widely used wireless technologies in smart agricultural solutions, based on FAHP results, yields the rankings shown in Figure 10. The wireless technologies referenced in the literature include WiFi, LoRa, Bluetooth, Zigbee, LTE, and Z-Wave. Additionally, a hierarchical assessment is conducted considering attributes such as cost, power consumption, range, delay, and capacity.

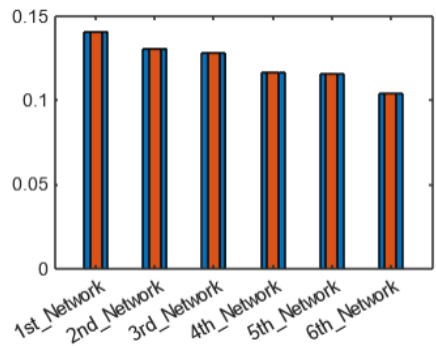

*Figure 10: Heterogeneous wireless networks ranking*

Assessing heterogeneous wireless networks using the Fuzzy Analytic Hierarchy Process (FAHP) yields a comprehensive ranking based on their performance metrics. Topping the list is the 1st Network, WiFi, with a score of 0.141, known for its robust data transmission capabilities and widespread use in various applications. Following closely in 2nd place is LoRa, which achieved a score of 0.131 and is recognized for its long-range connectivity and low power consumption, making it ideal for IoT applications. In 3rd place, Bluetooth scored 0.128 for its short-range communication and versatility on personal devices. Zigbee ranks 4th with a score of 0.117, valued for its energy efficiency and suitability in smart home environments. LTE, ranked 5th with a score of 0.116, is distinguished by its mobile broadband access and high-speed data services. Finally, Z-Wave ranks 6th with a score of 0.104, primarily favored in home automation for its reliable low-power wireless transmission. This ranking underscores the diverse strengths of each network in catering to specific technological needs.

5.2 Sensor Allocation for minimum length and maximum coverage

The agricultural area under examination is a deterministic zone measuring 300 meters by 300 meters, with a total surface area of 90,000 square meters. Within this defined space, sensors are strategically deployed in a uniform distribution, resulting in 36 sensor nodes arranged in a 6x6 grid. This systematic arrangement ensures comprehensive coverage of the entire area, as illustrated in Figure 11, highlighting the meticulous planning required for sensor deployment.

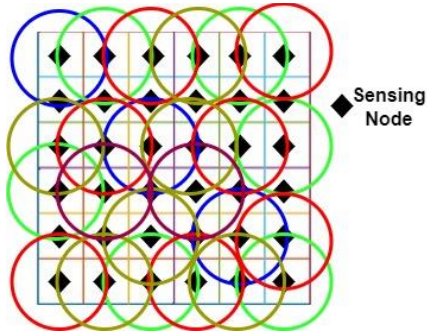

*Figure 11 Uniform distribution 36 nodes*

The deployment of 36 nodes incurs high implementation costs, consumes significant power, and introduces delays. However, it achieves the highest sensing coverage and offers highly efficient sensing of agricultural field parameters. Nonetheless, the high costs and power demands pose challenges to its practical implementation. An alternative approach employs the Fibonacci optimization method to select partial square grids. Those partial squares are chosen to distribute sensors effectively across the agricultural field, as shown in Figure 12.

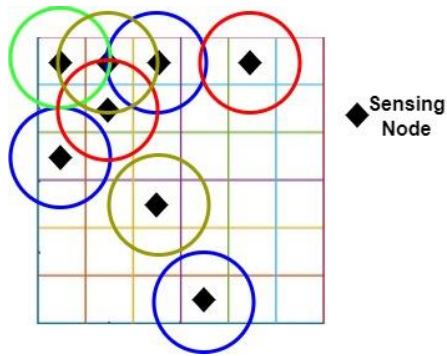

Figure 12: Fibonacci distribution of sensor units

As illustrated in Figure 6, the solution divided the area into a systematic grid of 36 square blocks, each 50 m on a side [14]. Only a few square blocks are occupied by sensor nodes, following the Fibonacci sequence: 1, 2, 3, 5, 8, 13, 21, and 34. This sensor distribution of sensor nodes follows the Fibonacci series given in equation (4). Fibonacci's selected-squares method has reduced the number of sensor nodes from 36 to 8. However, it lacks efficient sensing coverage of the agricultural area, covering only 45% of the total area and exhibiting a non-uniform distribution, creating more gaps between sensor nodes, as illustrated in Figure 6. On the other hand, using hybrid Gradient descent and Lagrangian optimization, and employing the Fibonacci function as the cost function to obtain the sensor node allocation distance $d_i$.

The Simulation for the gradient-based iteration with Lagrangian optimization assumed a side length of 300 meters and a minimum distance (min_distance), which helps keep sensors away from the boundaries. The initialization process randomly assigns sensor positions within the square area (0 to side_length for both x and y coordinates). The optimization process begins by listing the

Optimization Variables; each sensor has a 2D position [x, y]. The Objective Function (Simplified for Square) The calculate_objective_square function now encourages sensors to be closer to the center of the square (as a simplified proxy for covering the area) while penalizing the squared distance from the center (as a proxy for power) by keeping in mind the Fibonacci output as a total of eight sensor nodes. Then the gradients are now 2D vectors for each sensor, indicating the direction to move in the x-y plane to improve the objective. Along with considering the constraints and system boundaries:

- We enforce that the sensor positions stay within the square boundaries, with a minimum distance min_distance from each edge.
- Lagrange multipliers (lambda_xmin, lambda_xmax, lambda_ymin, lambda_ymax) are introduced for the four boundary constraints.
- The update rules for these multipliers push the sensors away from the boundaries if they get too close.

The Simulation yielded a shifted anchor node and a distance value multiplied by 165 within 500 iterations, as illustrated in Figure 13.

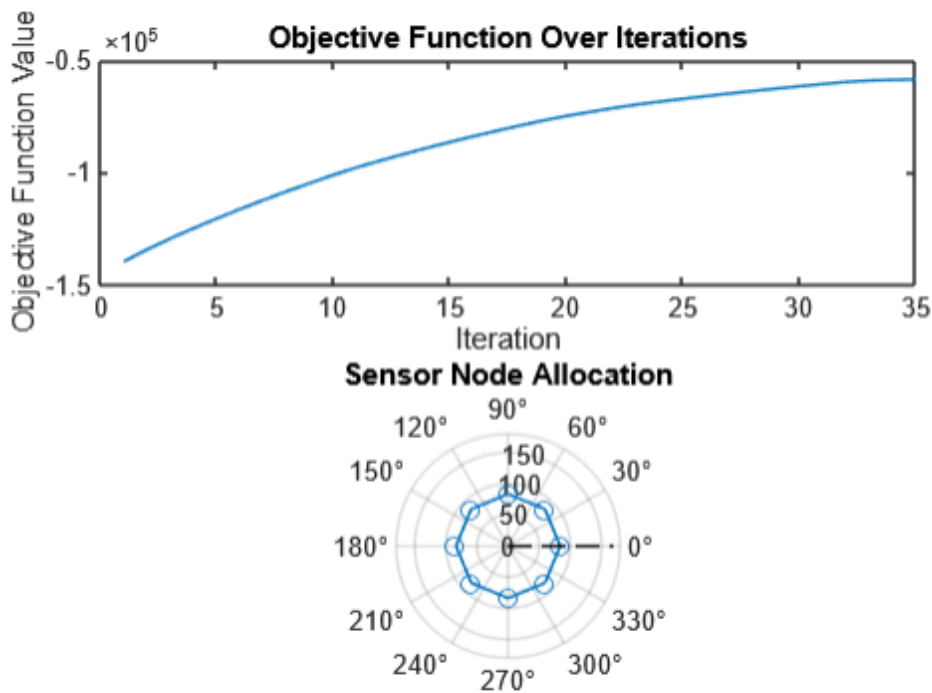

Figure 13: Inclined distance determination with the PSI angle

The gradient descent algorithm iteratively adjusts each sensor node's x- and y-coordinates within the square region. The objective function guides movement towards the center (for coverage) while minimizing distance from the center (for power). The Lagrange multipliers and the direct projection ensure that the sensors remain within the defined boundaries.

```
Optimization converged after 31 iterations.
Optimal distances from the central anchor (meters):
   80.3970
   80.3970
   80.3970
   80.3970
   80.3970
   80.3970
   80.3970
   80.3970
```

Figure 14: Maximum Distance result converged after 31 iterations

Aligning the sensor nodes within the square grid area Ac resulted in an equal-distance allocation equal to the Golden ratio, as shown in Figure 14. The proposed solution provides a better allocation of sensor nodes by spacing them at an interspacing of 1.618 cell units, corresponding to 80.9 meters. This allocation yields nine sensor nodes rather than eight. The extra node is the cornerstone of the allocation process, serving as the anchor or central node for the whole area, as illustrated in Figure 15.

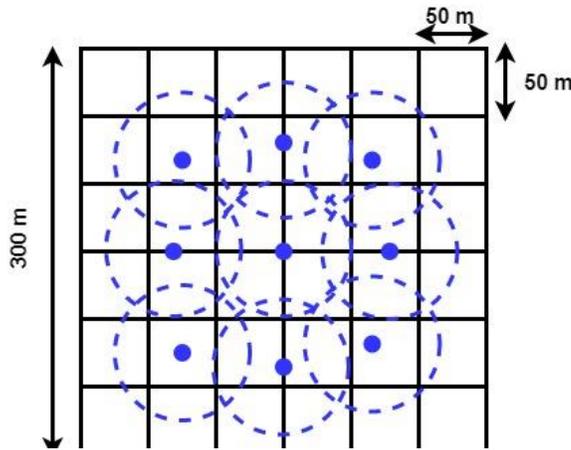

Figure 15 Circular deployment of the sensor nodes around the anchor node

As illustrated, the final deployment of the nine sensor nodes demonstrated impressive results when implementing the proposed hybrid algorithm, achieving 85% coverage across the total area and 100% within the sensing interest. This efficient arrangement allowed optimal monitoring and data collection across the designated area, significantly enhancing overall performance and reliability. However, this sensor distribution didn't account for scalability and reliability. As illustrated in Figure 15, the circular distribution provided a single point of failure. If the central anchor node fails, then all other nodes will fail as well. This dependence on a single point may also fail if too many sensing activities are sent to the same device simultaneously. Scalability is also an issue because all sensors depend on the scalability factor to communicate and report the sensing parameters. As more sensors connect in a star topology, the delay increases and the bandwidth decreases. Both communication parameters are significant for maintaining a robust WSN in smart agriculture systems. Considering those constraints, the Simulation maintains a redistribution of the sensor nodes.

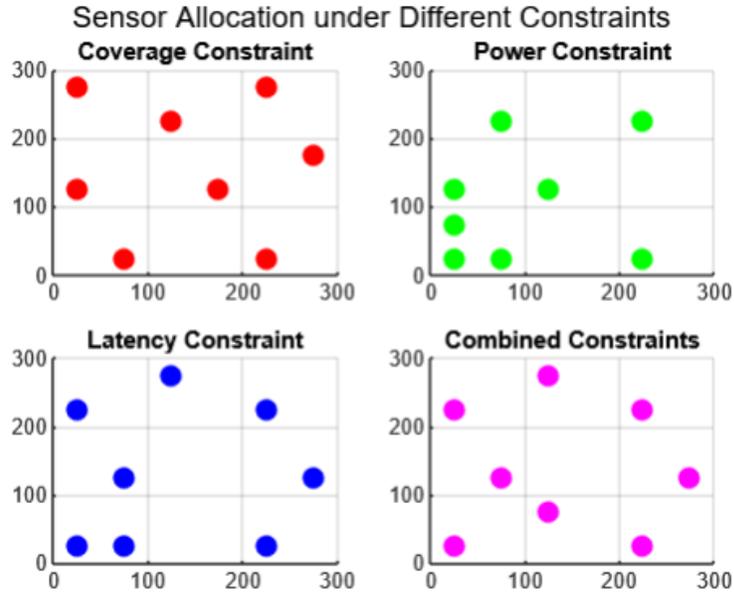

Figure 16 Sensor node allocation with maximum distance only

The sensor node distribution, subject to transmission constraints, yielded a new sensor allocation that requires improved alignment. The sensor allocation using the proposed method resulted in a sensor-node distribution with high overlap. Reforming the sensor nodes with a minimum overlap threshold of 10% and a maximum overlap threshold of 30% resulted in a better allocation, as illustrated in Figure 17.

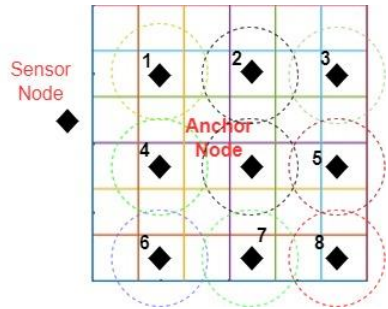

Figure 17: Alignment of the sensor node allocation based on maximum distance

The sensor nodes are distributed with an equal spacing of 1.6 square units, which is approximately equal to the golden ratio. This sensor node distribution achieved the objective of providing both improved sensing distribution and coverage. The coverage maintained 95% of the total area, considering better anchor node alignment. Moreover, the distribution showed better scalability, as it required fewer dependencies on the central node in a partial-mesh topology. The distribution was maintained to minimize latency, maximize coverage, and minimize power consumption. Moreover, the distribution maintained better coverage with minimum and maximum overlap thresholds of 10% and 30%, respectively.

### 5.2.1 Extra node distribution for more sensing coverage

Further investigation considers the scalability of the sensor nodes' distribution to accommodate additional specific and Constraint sensors. Those sensors are effective at observing additional areas or objects via Bluetooth, a wireless technology. This is investigated further in MATLAB to extend the sensing uncertainty by employing suboptimal extra-sensing localization via gradient descent and Lagrangian methods. The total area remains the same at 90,000 square meters. The anchor node position is the fixed central position. The number of Wi-Fi sensor stations is 8, as previously determined. The number of additional Bluetooth sensor nodes to be deployed will be set by initial_extra_node_count (e.g., 100, given the maximum capacity per network). This value is fixed; the gradient descent optimizes their *positions*, not their *count*. Ranges confirm the configured WiFi and Bluetooth ranges. The Gradient Descent Parameters reports the learning_rate, num_iterations, and the strengths of the different forces (repulsion_strength, attraction_strength, lagrangian_penalty_strength). These parameters are crucial for tuning the optimization process. If num_iterations is low or learning_rate is too small, the nodes might not fully converge. If the learning rate is too high, the process might become unstable.

Assumptions for the MATLAB Simulation:

- WiFi Range (Anchor to Stations): 70 meters (configurable).
- Bluetooth Range (Station to Extra Nodes): 15 meters (configurable).
- Extra Node Placement Strategy: We'll use a grid-based approach, placing extra sensor nodes at regular intervals that radiate outward from the initial eight sensor stations. Each additional node will attempt to connect to the nearest sensor station within Bluetooth range, or to an already connected Bluetooth node.
- Coverage Visualization: Coverage will be represented as circles with their respective ranges, assuming an ideal omnidirectional antenna.

The simulation results are shown in Figure 10.

```
Starting Gradient Descent...
Gradient Descent Finished.

--- Network Statistics (Gradient Descent) ---
Total Area: 90000.00 sq meters
Anchor Node Position: (150.00, 150.00)
Number of Wi-Fi Sensor Stations: 8
Number of Extra Bluetooth Sensor Nodes Deployed: 100
Wi-Fi Range (Anchor to Station): 70.00 meters
Bluetooth Range (Station/Node to Node): 15.00 meters
Gradient Descent Learning Rate: 0.500
Number of Iterations Completed: 299
Repulsion Strength: 0.50
Attraction Strength: 0.80
Lagrangian Penalty Strength: 10.00
>>
```

Figure 18 Simulation results for allocating extra sensors

The simulation iteration resulted in 100 additional, more constrained sensors that use Bluetooth wireless technology to connect to one of the eight sensor stations. However, more illustrations of the extra-formed sensor nodes are shown in Figure 19.

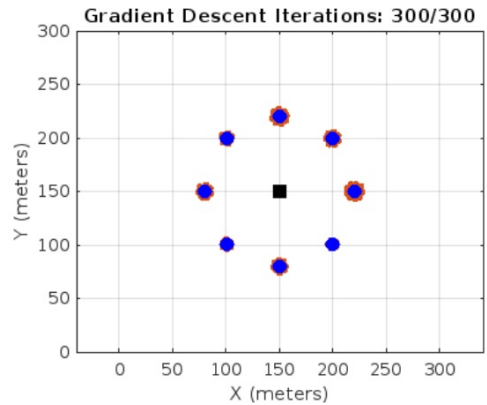

Figure 19: Extra node allocation

This Figure animates the movement of the "Extra Sensor Nodes (Bluetooth)" during the gradient descent iterations. The nodes initially cluster somewhat around the eight blue sensor stations because they were initialized at random within their assigned station's range. One can observe the orange-dotted nodes (additional Bluetooth sensors) gradually spreading across the simulation area. They are:

- o Repelled from each other: This prevents excessive clustering and encourages them to cover more distinct areas.
- o Attracted to their assigned blue sensor stations: This ensures they remain within connectivity range (or at least strive to).
- o Pushed away from boundaries: This keeps them within the defined 300m x 300m square.
- o Pulled towards their station by Lagrangian penalty: If a node drifts too far from its assigned station, the Lagrangian multiplier (lambda_conn) for that node increases, applying a stronger "pull" back towards the station and enforcing the connectivity constraint.

After num_iterations, the nodes should have settled into a more distributed pattern, aiming to balance coverage and connectivity. The animation gives a more visual sense of the optimization process. Figure 11 illustrates the optimization of sensor deployment for WiFi and Bluetooth connectivity.

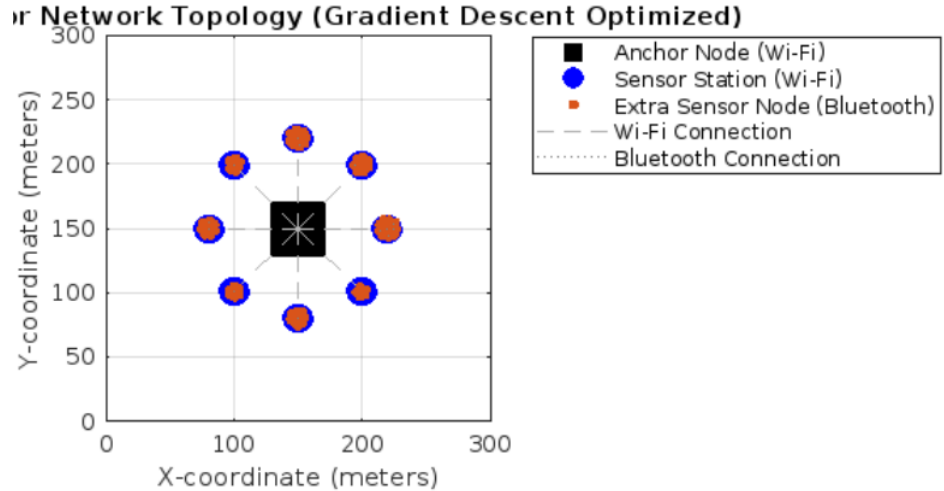

Figure 19: Extra node wireless communication

This Figure presents the final, optimized layout of the entire sensor network after the gradient descent algorithm has completed its iterations. Black Square (Anchor Node) remains at the center (150, 150) and serves as the main hub. Blue Circles (WiFi Sensor Stations) remain positioned symmetrically around the anchor, connected by dashed gray lines indicating their WiFi links to the anchor. These serve as the base for the Bluetooth expansion. Orange Dots (Extra Sensor Nodes - Bluetooth): These are the nodes whose positions were optimized. One can notice they are:

- o More evenly spread: Compared to a purely random or simple grid, the gradient descent aims to distribute them more uniformly, reducing large gaps and excessive overlaps.
- o Connected to Stations: The dotted gray lines show their Bluetooth connections. Due to the attraction and Lagrangian penalty terms in the optimization, these nodes are generally within Bluetooth range of their assigned sensor station. This commitment to range distribution ensures the expanded network remains connected.
- o Covering the "remaining area": They fill in the space between the initial eight sensor stations and extend outwards towards the edges of the 300m x 300m square, effectively increasing the overall sensing range.

More illustrations of coverage and wireless communication are shown in Figure 12.

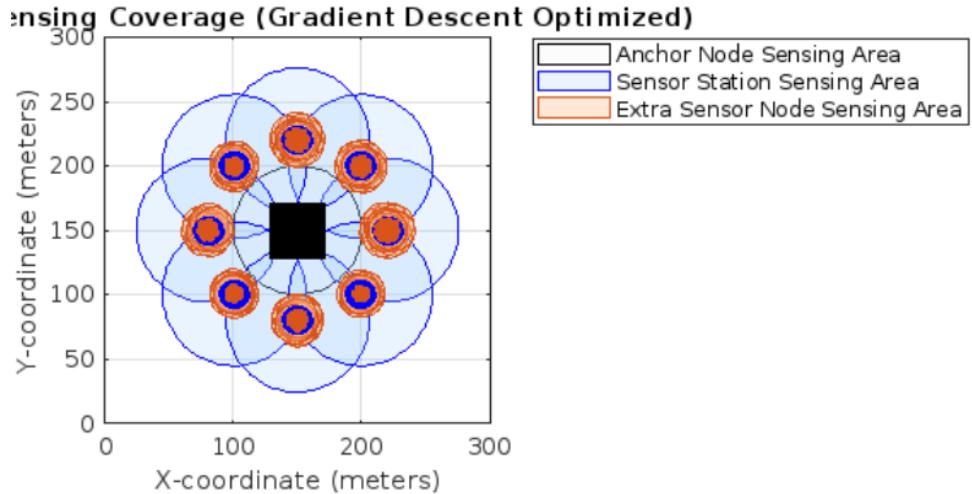

Figure 20: Extra sensor allocation with wireless coverage

This Figure visually depicts the approximate sensing range of each node type, providing a clear picture of the network's total coverage area. Light Gray Circle (Anchor Node Sensing Area): A symbolic representation of the anchor's immediate vicinity or centralized data collection capability. Light Blue Circles (Sensor Station Sensing Area): The eight larger blue circles indicate the sensing range of the WiFi sensor stations. These provide the initial layer of broader coverage. Light Orange Circles (Extra Sensor Node Sensing Area): These are the most critical for understanding the "increased sensing range." Each orange circle represents the sensing coverage of an individual Bluetooth node.

- o Overlap: There is a significant overlap between these orange circles. This overlap is desirable in sensor networks for redundancy, robustness (if one node fails, nearby nodes still cover the area), and improved data accuracy through multiple readings.
- o Area Filling: The key result here is how these orange circles fill in the gaps left by the blue sensor station circles and extend coverage further out into the 300m x 300m area. The gradient descent has pushed them into positions that maximize this spread while maintaining connectivity.
- o Boundary Coverage: The combined effect of node repulsion and boundary repulsion often results in a high node density near the edges of the simulation area, ensuring that the entire square is sensed.

Finally, by optimally allocating and aligning both the sensing station and the extra Bluetooth sensors around the anchor node, as in Figure 17, the Final distribution for nodes with both WiFi and Bluetooth is illustrated in Figure 21.

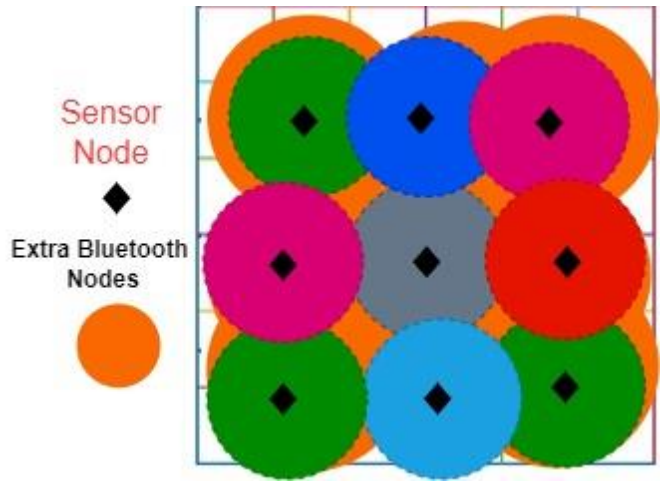

Figure 21 Final heterogeneous sensor node distribution

Gradient descent with a Lagrangian-like penalty optimally allocates additional Bluetooth sensor nodes to increase the sensing range.

- Optimality: In this context, "optimal" means finding a node distribution that effectively spreads out the sensing coverage (due to repulsion) while strictly adhering to the connectivity requirements (due to attraction and the Lagrangian penalty). It provides a more organic, efficient distribution than a rigid grid, adapting to the initial placement of the WiFi stations.
- Increased Sensing Range: The combination of the central anchor, the 8 WiFi sensor stations, and the numerous Bluetooth extra nodes creates a multi-layered sensing network. Bluetooth nodes significantly extend coverage, enabling monitoring of areas beyond the reach of the initial Wi-Fi stations alone, thereby increasing the system's overall sensing range within the 300m x 300m area.
- Self-Organization (Simulated): This approach mimics self-organization in which nodes "decide" on their best positions based on local interactions (repulsion, attraction, and connectivity rules).

This Simulation demonstrates a powerful computational method for optimizing sensor network deployments, particularly in scenarios where dynamic adjustments or complex objectives (e.g., balancing coverage, connectivity, and energy) are required.

**5.2 Comparing the proposed hybrid algorithm with the solutions found in the literature**

Figure 22 compares the proposed hybrid algorithm with the systematic uniform and Fibonacci distributions. This highlights the innovative efficacy and performance enhancements of our approach compared to established methods.

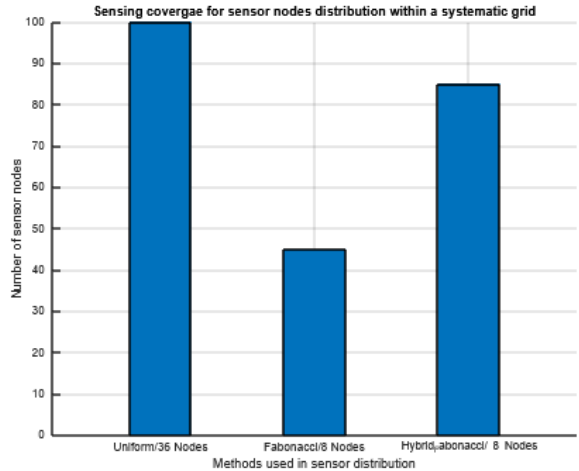

Figure 22: Comparing the proposed sensor allocation based on coverage

The evaluation of various sensor node distribution methods within a systematic grid allocation reveals intriguing insights into their effectiveness in achieving sensing coverage. Specifically, the uniform distribution method delivers 100% sensing coverage, ensuring that every area is monitored without gaps. In contrast, the Fibonacci distribution method falls short, achieving only 45% sensing coverage, thereby significantly limiting its effectiveness for comprehensive environmental monitoring. However, the proposed hybrid algorithm stands out by striking a remarkable balance, providing 85% sensing coverage while optimizing the sensor nodes' placement. This innovative hybrid approach not only enhances coverage but also proves to be the most efficient in terms of coverage per sensor node.

Additionally, from an economic perspective, the cost of sensor allocation using the proposed hybrid algorithm is substantially lower than that of the uniform distribution method. This highlights its practicality for real-world applications where budget constraints are critical. The combination of reduced costs, lower power consumption, and superior coverage-to-node ratio underscores the effectiveness and advantages of the proposed hybrid algorithm in sensor distribution strategies.

**Comparing the proposed hybrid algorithm sensor distribution with deterministic methods**

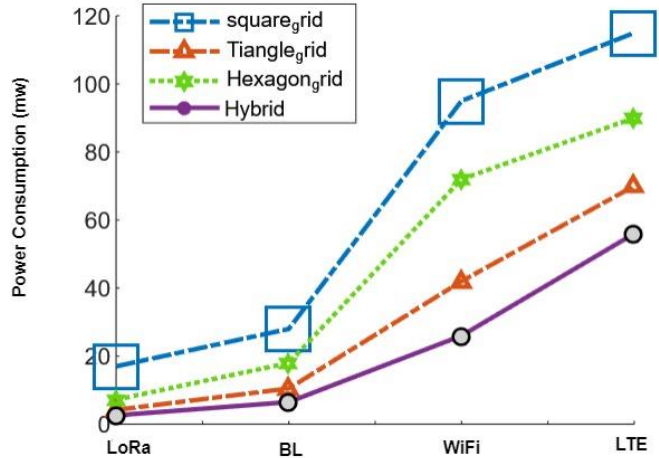

Figure 23: Comparison of the proposed hybrid method in sensor allocation versus the deterministic methods

The proposed hybrid algorithm achieved a lower power consumption rate than the classic deterministic sensor node distribution reported in the literature, as shown in Figure 23. The consumption rate is notable, particularly for Wi-Fi, which is widely used as the primary wireless technology for sensor connectivity. The proposed heterogeneous connectivity achieved high, sustainable connectivity rates by ensuring heterogeneity. This heterogeneity addresses the connectivity interoperability issue, which is a challenging problem for industrial sensors, particularly in smart agriculture systems.

**Comparing the proposed hybrid algorithm sensor distribution against the PSO**

The proposed hybrid algorithm is compared with PSO, the most cited and well-known metaheuristic in the literature. The comparison depends on the main result of both algorithms, with the same considerations but without alignment.

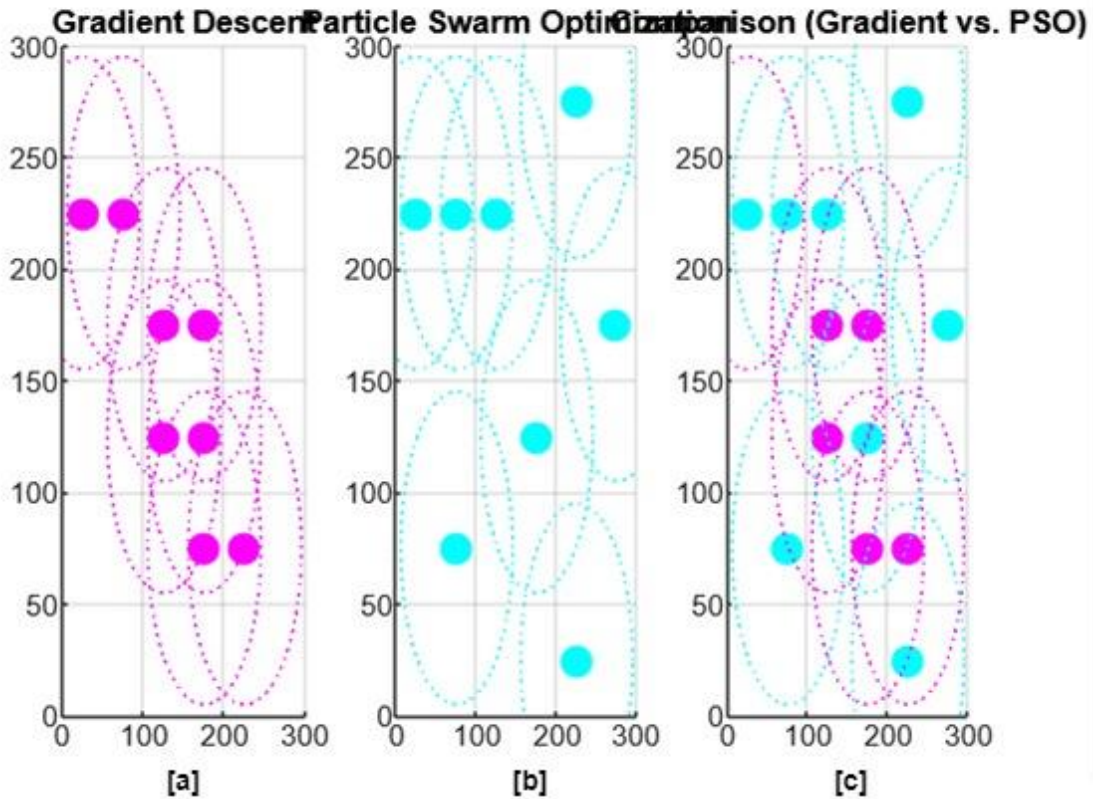

Figure 24: Comparing the proposed hybrid method in sensor allocation with the PSO method

The proposed hybrid algorithm, compared with particle swarm optimization (PSO), provided better wireless coverage, as illustrated in Figure 24. The proposed hybrid algorithm distribution is illustrated in Figure 24. a provides a better sensor distribution across the agricultural cluster area, as all sensors converge to improve sensing and wireless coverage. The PSO distribution of sensor nodes illustrated in Figure 24.b produced more distributions with minimum sensing coverage because it diverged and left gaps in sensing coverage. Figure 24.c illustrates the sensor distribution of the proposed hybrid algorithm, which exceeds that of the PSO.

## Conclusion

The paper investigated the implementation of sustainable connectivity for the smart agriculture system to ensure real-time monitoring of agricultural processes. The paper achieved the objective in two ways: first, by providing a heterogeneous wireless infrastructure. The FAHP was employed to rank the most applicable wireless technologies within the sensor networks. The second contribution was achieved by providing a distribution of sensor-node distances that maximizes sensing range while minimizing the number of sensors. Gradient descent with Lagrange was used to find the maximum possible distance for an inclined psi angle. The proposed solution provided a circular and square grid that minimized delay and power consumption and maximized coverage. The methodology constructs the solution by treating the Fibonacci sequence as the objective function and aligning the sensor distribution with equal spacing at the golden ratio of 1.6. The research was compared with both the classic deterministic and PSO distributions and demonstrated its superiority. As future work, the proposed approach should be evaluated through real-world empirical testing to assess sensor deployment error and accuracy.